# UC Santa Cruz
## Activity Descriptions

**Title**

Galaxy-Classification Activity for All Ages

**Permalink**

https://escholarship.org/uc/item/2tk5j8zh

**Authors**

Cooksey, Kathy L
Metevier, Anne J
Rubin, Kate HR
et al.

**Publication Date**

2022-09-19

**Copyright Information**

This work is made available under the terms of a Creative Commons Attribution License, availalbe at https://creativecommons.org/licenses/by/4.0/

# Galaxy-Classification Activity for All Ages

Kathy L. Cooksey[*1], Anne J. Metevier[2,3], Kate H. R. Rubin[4], Philip I. Choi[5], and Lynne Raschke[6]

[1] Department of Physics & Astronomy, University of Hawai'i at Hilo, Hilo, HI, USA
[2] Department of Physics & Astronomy, Sonoma State University, Rohnert Park, CA, USA
[3] Department of Earth & Space Sciences, Santa Rosa Junior College, Santa Rosa, CA, USA
[4] Department of Astronomy, San Diego State University, San Diego, CA, USA
[5] Department of Physics & Astronomy, Pomona College, Claremont, CA, USA
[6] Department of Mathematics & Physics, The College of St. Scholastica, Duluth, MN, USA
[*] Corresponding author, kcooksey@hawaii.edu

## Abstract

Classification is a general tool of science; it is used to sort and categorize biological organisms, chemical elements, astronomical objects, and many other things. In scientific classification, taxonomy often reflects shared physical properties that, in turn, may indicate shared origins and/or evolution. A "hands-on" galaxy-classification activity developed and implemented by Professional Development Program (PDP) participants, for a high-school summer STEM enrichment program, has been adopted for various age groups and venues, from young (K–3) to college students. We detail the basic tools required, outline the general activity, and describe the modifications to the activity based on learners' ages and learning objectives. We describe the facilitation strategies learned through PDP training and used when implementing the activity, including prompts to motivate the students. We also discuss how we connected the classification process to astronomy and science more broadly during the concluding remarks.

Keywords: activity design, astronomy, classification, facilitation, galaxies

## 1. Introduction

"Hands-on" learning activities fall along a spectrum, from guided worksheets to open-ended exploration, and the choice of approach ought to align with the learning objectives (LOs) and logistical constraints (Institute for Inquiry [IfI], 2006a; Rice, 2010). Some topics might be too abstract for a hands-on approach, and galaxies (and most astronomical objects) definitely do not physically fit in the classroom! However, a main tool of astronomy is observations, and images are possible for students to lay hands on, as well as being a rich resource.

The standard galaxy classification scheme is based on morphology (overall shape and smaller-scale structure), but other physical properties, primarily color, also distinguish morphologically classified galaxies. In addition, galaxies within a class underwent a relatively self-similar evolution. Thus, the hands-on galaxy-classification activity described here opens the door to learning about galaxies, their constituent parts, galaxy evolution, and properties of observational astronomy (e.g., light, color, imaging).

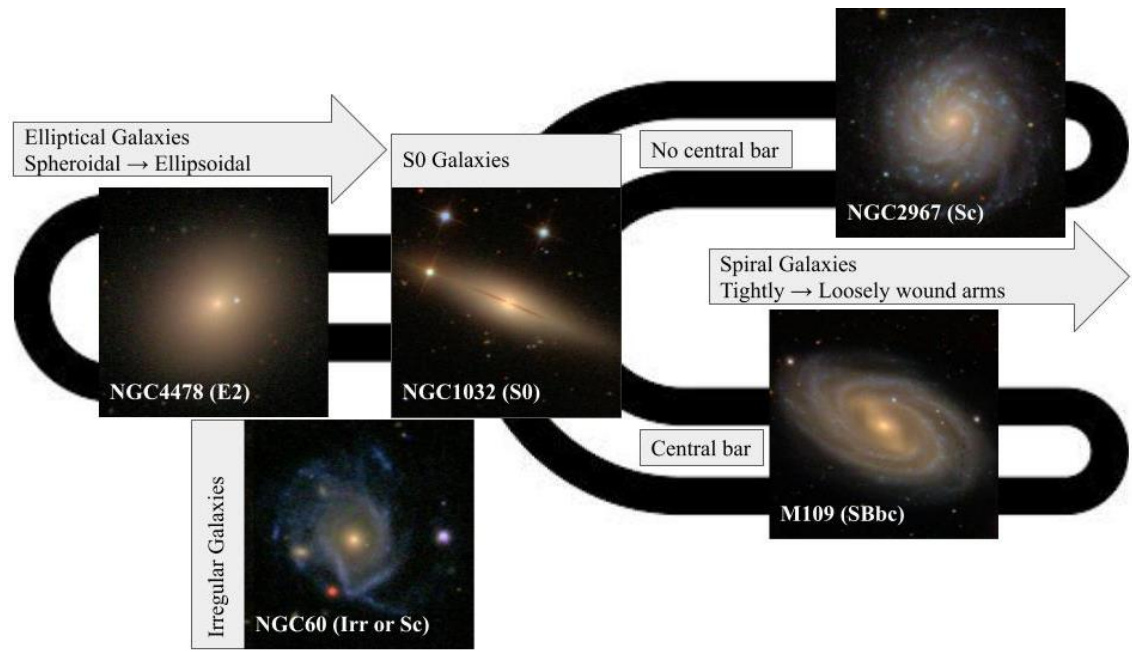

**Figure 1: Schematic of the "Hubble tuning fork," a morphological classification of galaxies.** Images are examples from the suite used in the galaxy-classification activity (see §2), from the Sloan Digital Sky Survey (Alam et al., 2015). They are labeled with the New General Catalogue (NGC) or Messier (M) identifier, and, parenthetically, the accepted classification: elliptical (E) with 0–7 subclass; S0 also known as lenticular; unbarred spiral (S) or barred spiral (SB), both with subclassification a–c; or irregular (Irr). NGC60, here placed as an irregular galaxy, is visually classified as either an irregular or unbarred spiral; it demonstrates there is modest subjectivity in visual classification.

Moreover, taxonomy is general across science disciplines and can lead learners to question or hypothesize about why objects end up in different classes, where asking questions and hypothesizing are science process skills (IfI, 2006b). By classifying objects, learners engage in an authentic science experience that, ideally, promotes interest in and motivation for the pursuit of science. Color images of galaxies are fun to look at (see Figure 1), and a carefully selected suite of galaxy images can facilitate learners "correctly" classifying the galaxies, though "correct" is not necessary and possibly misleading. First, professional astronomers would possibly disagree on the exact classification of many galaxies.

Second — and the real learning kicker, astronomical images are two-dimensional renderings of three-dimensional objects. Thus, there are projection effects that must be considered; for example, a disk galaxy (think frisbee) seen face-on would be a circle but more "cigar" shaped edge-on (compare NGC2967 to NGC1032 in Figure 1). Since we only have our Earth-centric view of the Universe, astronomers must study many self-similar systems across the sky to understand their 3D structure. This nuance enables exploration and discussion of the scientific process of astronomy.

The Professional Development Program (PDP) trained scientists, engineers, and educators to teach through inquiry, to engage learners equitably and inclusively, and to assess learning gains. "Inquiry" is teaching science as science is done, with learner-driven, iterative, facilitated investigation (see Metevier et al., 2022, and Metevier et al., this volume, for a discussion of inquiry and authentic, inclusive STEM learning experiences). The Center for Adaptive Optics (CfAO), an NSF Science Center at the University of Santa Cruz (UCSC), originally developed and ran the PDP, which was later run by the Institute for Scientist & Engineer Educators (ISEE); the PDP ran from 2001–2020. It was a training ground and sandbox for designing and redesigning inquiry-oriented activities and for developing and practicing facilitation techniques. The galaxy-classification activity described in this article was developed by PDP participants for an affiliated teaching venue.

This galaxy-classification activity can be used as an inquiry "starter" or a stand-alone activity, for different educational venues. A "starter" is an introductory activity; it could be a demonstration of a phenomenon and encouragement of learners to raise questions or an activity that involves learners in the processes of science and piques their interest in the rest of the upcoming learning experience (for more on "starters," see Kluger-Bell, 2010). We begin with a brief overview of galaxy properties relevant to the activity in §1.1. The activity itself, including alternative implementations based on learner age, is described in §2. Social and cultural aspects, including those for learners with disabilities, are briefly broached in §3. We also discuss, in §4, how the activity (or elements thereof) can be used as a precursor to more investigation.

First, we briefly discuss terminology. Aligning with ISEE custom, we will often refer to "learner" and "facilitator" instead of "student" and "instructor," respectively, in the sections below. "Learner" keeps the discussion general to different educational venues, from classrooms to public-outreach events. "Facilitator" emphasizes the educator is respecting the learners' approach, formatively assessing where the learners are at with respect to the LOs of the activity, and encouraging them toward those objectives (see Kluger-Bell et al., this volume). Facilitation is closely linked to the three ISEE themes of inquiry, equity and inclusion, and assessment. There are various techniques for successful facilitation, and, naturally, a facilitator's style should be considered when choosing what approach to use. The participant's sense of ownership of the learning is an overarching consideration for inquiry-activity designs. Thus, we emphasize facilitation strategies to foster and respect ownership (also see Ball et al., 2022).

## 1.1. Brief overview of galaxies

To provide a common, bare-bones framework for discussing galaxies, we briefly summarize relevant points about galaxies, including introducing astronomical jargon. Readers familiar with the topic may skip this section.

*Classification:* Galaxies are primarily composed of stars, gas, dust, black holes, and dark matter; they come in a range of shapes, sizes, and colors. Figure 1 illustrates the "Hubble tuning fork" with images used in the classification activity described in §2. The "Hubble tuning fork" classification places increasingly elliptical galaxies along the stem and more loosely wound spiral (aka disk) galaxies along the tines (with one tine having spirals with a bar in the center and unbarred spirals along the other tine); at the shoulder of the tuning fork are S0 galaxies, which have properties of both elongated elliptical and disky spiral galaxies. In this canonical framework, irregular-shaped galaxies are lumped off to the side.

*Morphologies:* The stars, gas, and dust in spiral galaxies are distributed like flat frisbee disks; they are "spiral" because the stars, gas, and dust are concentrated in pinwheel-like arms. Some spiral galaxies are very flat, when seen edge-on (inclination of 90°). Others have a central bulge and look like a ball wedged inside a donut; seen edge-on, these spiral galaxies look like NGC1032 in Figure 1. All spiral galaxies are fairly round when seen face-on (inclination of 0°). Thus, imaging studies of spiral galaxies are highly subject to 2D projection effects, depending on the galaxies' angles with respect to Earth; in Figure 1, NGC2967 is face-on, M109 is inclined 66°, and NGC1032 is edge-on. Elliptical galaxies are overall round, from spheroidal to ellipsoidal (think rugby ball); they are less subject to projection effects. Irregular galaxies, as the name implies, typically have unusual 3D shapes, which may result in 2D projection issues.

Spiral and irregular galaxies tend to appear "clumpy," since their constituent stars and gas are not distributed uniformly. Elliptical galaxies, on the other hand, are "smooth" — more uniform light distributions with little substructure. S0 galaxies have the shape of a disk galaxy but are smooth like an elliptical galaxy. Nearly all galaxies have a supermassive black hole and a high concentration of older stars at their centers; thus, galaxy centers look relatively similar: bright, yellow-white, and smooth.

*Colors*:[1] A young, recently formed single population (grouping) of stars is overwhelmingly blue because the brightest young stars are hot and blue. After a long time, a single stellar population becomes redder because the long-lived stars are cool and red. However, when stars (of whatever color) are viewed through a lot of dust, the light reddens (think of sunsets or sunlight through fire smoke). Spiral galaxies are actively forming stars from gas and dust, and, due to their disk morphology, tend to be blue when viewed more face-on and redder when viewed edge-on (compare face-on NGC2967 to M109 inclined 66° in Figure 1). Due to the evolutionary effects discussed below, elliptical galaxies are composed of mostly old stars and tend to be red, with reduced signatures of gas and dust. Star-forming irregular galaxies lean toward the color properties of spiral galaxies: blue unless reddened by dust.

*Evolution*: At the most cursory level, the Universe evolved to form spiral galaxies first (which may have looked irregular as they formed; see Governato et al., 2008). When two or more spiral galaxies merge, they become irregular for a period. If one galaxy is sufficiently larger than its merger partner(s), the new system may relax into a larger spiral galaxy. The final outcome of comparable-sized spiral galaxies merging is, typically, an elliptical galaxy (see Jonsson et al., 2008). Thus, galaxy origin/evolution is reflected in the "Hubble tuning fork" classification.

# 2. Activity description

## 2.1 Original design of the activity "starter"

The original galaxy-classification activity was developed for the California State Summer School for Mathematics and Science (COSMOS), in 2002. COSMOS is a month-long residential academic and enrichment experience for high-school students; UCSC is one of the four UC campuses that host a program.[2] The CfAO PDP, which was later run by ISEE, ran an astronomy course through COSMOS from 2001 to 2007; it is fully described in Cooksey et al. (2010, and references therein). COSMOS participants were organized into topical "clusters" of 16–18 students, where the CfAO-led cluster historically combined astronomy and vision science, due to the overlapping use of adaptive optics. All COSMOS students participate in a research project in their cluster and share their results on the last day of the program. The CfAO-led cluster designed two-week-long (roughly 20–30 hours), inquiry "research" projects for groups of two to three students with one dedicated project advisor. The projects were designed to feel authentic to the learners and meet the LOs of the program.

For the Galaxy Morphologies project, the classification activity was used as a "starter" for the research investigation. The group was given a suite of color galaxy images and prompted to classify them; they were facilitated to consider the shapes and colors (see §2.1.1). At this point in the COSMOS astronomy course, they had a brief introduction to the "Hubble tuning fork" (see Figure 1), so there was a seed for the group to gravitate towards spiral, elliptical, and irregular galaxies. Yet the suite of images was selected to have ambiguous cases (e.g., NGC60 in Figure 1) so the group could discuss the properties and the group plus advisor could discuss the subjective nature of visual classification.

---

[1] We refer to visible (optical) light, so colors are what the human eye would perceive.
[2] See https://cosmos.ucsc.edu.

There can be two perspectives on the LOs of an educational design: the learners' and the facilitators'. Often, the learners are content-oriented and engage with the activity to learn about concepts relating to a particular topic (e.g., astronomy); of course, the facilitators want learners to understand new content, too. In addition, the facilitators may have LOs that are process- and/or motivation-oriented. "Process" refers to learners engaging in (and possibly improving in) the practices of science: observing, inferring, interpreting, etc. (Ifl, 2006b). This experience can be empowering for the learners, and the facilitators may draw attention to how the learners were scientists and motivate them to pursue science courses and careers. The LOs for the COSMOS classification "starter" include:

- Learners are capable of hypothesizing a classification scheme with equal legitimacy as the "Hubble tuning fork" as supported by their justification of their classes.

- Learners use correct terminology for galaxy morphologies and correctly identify trends between shapes and colors; evidence includes the learners understanding galaxy taxonomy and its relationship to other galaxy properties.

- Learners show they understand projection effects by, for example, articulating how e.g., spiral galaxies seen edge- versus face-on are actually in the same class. "Projection effects" refer to the fact that astronomers use 2D images to understand our 3D Universe and only have a perspective from Earth.

The only equipment needed for this activity is a diverse suite of color galaxy images.[3] The bulleted activity overview is:

- Introduction: The project advisor shares a few images of galaxies; prompts the group to classify the suite of galaxies, being sure to justify their classification scheme; and specifies that there will be an informal share-out of the results.

- Activity time: The group works while the project advisor facilitates, as necessary, to achieve the LOs and handle any social dynamics or other issues. This takes roughly 25 minutes.

- Share-out: The group presents their classifications and explains their rationale. If not voluntarily addressed, the project advisor asks about any disagreements or residual questions.

- Synthesis: The project advisor leverages what the group did during activity time and presented in the share-out to highlight the LOs. The synthesis can be tailored to the students' observations and interests.

- Planning for next steps: The advisor facilitates the learners hypothesizing why galaxies have different shapes and colors, and they discuss plans for future investigation. The project advisor is transitioning the group to the remainder of the Galaxy Morphologies project, in which they select a galaxy to investigate; this includes retrieving images of the galaxy in multiple colors, researching its properties in the literature, and presenting their findings to other COSMOS students and instructors.

### 2.1.1 Facilitation strategies

The PDP trained participants in facilitation (see Kluger-Bell et al., this volume). Sound facilitation practices include respecting the learners' approach, formatively assessing where the learners are at with respect to the LOs of the activity, and encouraging them toward those objectives. In the PDP, it was of utmost importance to facilitate in such a way as to foster and respect ownership (e.g., supporting learners' approaches to the activity; also see Ball et al., 2022).

---

[3] Contact the corresponding author for her suite of images (examples in Figure 1), retrieved from the Sloan Digital Sky Survey (Alam et al., 2015) in 2015, from one of their learning modules like the following: http://voyages.sdss.org/expeditions/expedition-to-galaxies/galaxies-3/.

The facilitator of this activity (i.e., the project advisor) observed learners as they worked and asked questions about what they were working on, to gauge their progress with respect to the LOs. Specifically, the facilitator might look for learners using correct terminology to describe galaxy morphologies, being able to justify and explain their classification schemes, and demonstrating understanding of projection effects. As learners worked, the facilitator would make gentle course corrections if learners were not moving toward the LOs or ask probing questions to help them reach the LOs. For example, if two learners classified the same galaxy differently, the facilitator may have prodded: "I see that one of you classified this galaxy as an elliptical, and the other classified this galaxy as an S0. Why do you disagree?" This could lead to a rich, learner-led discussion of projection effects.

The facilitator also managed group dynamics, encouraging learners to work together and stay engaged in the activity. If one learner seemed to be hanging back, the facilitator could ask questions of them specifically, to give them an opportunity to demonstrate their understanding. The facilitator could also remind learners who took a more dominant role to make sure the other group members had opportunities to sort the images and describe their ideas about classification schemes.

## 2.2 Implementation for K–12 learners

The galaxy-classification "starter" has been adapted as a stand-alone activity for K–12 outreach (primarily K–6); a professional astronomer brought the activity to the classroom and acted as lead facilitator. The implementation described here consists of the galaxy-classification "starter" and a short lecture afterwards, with no extended follow-up investigation.

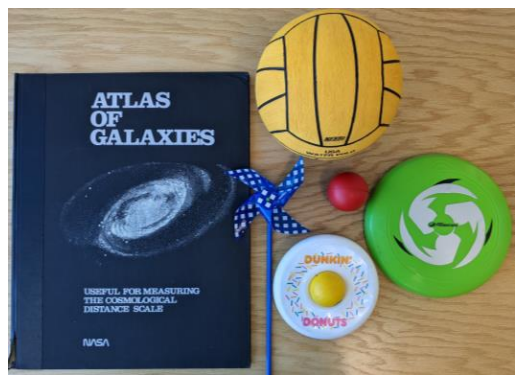

**Figure 2: Selection of equipment for the galaxy-classification activity.** The *Atlas of Galaxies* (Sandage & Bedke, 1998) is physically large and very visible; it can be used in the "starter" (see §2.2.1) to show black-and-white images of spiral galaxies. The other objects are used as physical models of galaxies in the concluding remarks (see §2.2.3), to demonstrate the 3D shapes and 2D projection effects of the galaxy types. Also useful but not shown is a rugby ball or American football.

As an outreach activity for younger learners, the primary LOs are motivation-oriented:

- Learners positively engage in an authentic science experience as demonstrated by the enthusiasm for active participation.

- Learners feel empowered to be interested in and, ideally, pursue science as demonstrated by the questions asked during the concluding remarks.

With the motivation focus, attentive facilitation became crucial to success.

There are content LOs such as "Learners understand classification is a general tool of science" and, nominally, the LOs of the original COSMOS activity (see §2.1). A useful LO to possibly add is about astronomical imaging (e.g., foreground and background objects, artifacts like bad pixels as seen in M109 in Figure 1, and use of filters to recreate color images) because the process of science, especially observing and "seeing through noise," is important to convey to young learners.

In younger classrooms, there are typically 20–30 students, one teacher, and one to two teaching assistants to help facilitate the activity as non-experts; in older classrooms, there are usually no assistants. Seventh- through 12th-grade students may be better served — and more engaged — by the college-level implementation described in §2.3, if not the original activity described in §2.1; the facilitation approach and details in the concluding remarks should be updated appropriately.

A typical class period is about 50 minutes. Assuming the classification activity is one period, the schedule roughly breaks down as: 15 min for introductions and instructions; 15 min for groups to classify the galaxies; 10 min for the share-out; and 10 min for concluding remarks. Equipment includes:

- Color images (one per learner), labeled with galaxy identifier and image source; the ensemble used by the class includes multiple examples of each galaxy type

- "Starter" images (e.g., big book of galaxies *a la* Sandage & Bedke, 1998, shown in Figure 2, *Hubble* Deep Field enlargement, or diverse galaxy group like Hickson Compact Group 44)

- Final handout with illustration of "Hubble tuning fork" (more elaborate than Figure 1) and a table of all galaxies and their "official" classification plus any additional suitable information (e.g., other common identifiers, coordinates)

- Physical models to illustrate galaxy structure: pinwheel, frisbee disk, donut-shaped disk that can fit a ball, balls of various sizes, and football or rugby ball (see Figure 2); useful to have the colors of the objects reinforce the colors of galaxies and/or their components

(e.g., a blue pinwheel, blue disks, yellow or red balls)

- Wide-angle light source (powerful lamp or overhead projector) where physical objects can be shown as shadows to demonstrate 3D-to-2D projection effects and viewing angle

- Equipment to project videos (e.g., Governato et al., 2008; Jonsson et al., 2008) and, if possible, demonstrate websites (e.g., Galaxy Zoo: https://www.zooniverse.org/projects/zookeeper/galaxy-zoo/)

Facilitating strategies for the activity are embedded in the following subsections (§§2.2.1–2.2.3).

## 2.2.1 "Starter," instructions, and active classification time

After introductions are made, the lead facilitator primes the class for the activity — and makes a formative assessment — by asking what they know about galaxies. The shared responses usually cover the basics that galaxies are big and made of stars and our Galaxy is the Milky Way, which we can also see at night; if not, the facilitator can elicit the basics (e.g., "what are galaxies made of?", "what galaxy do we live in?"). The facilitator respects all contributions; this ranges from thanking the learner for sharing to verbally requesting they remember to bring it up later since it is important. It is useful to clarify what learners say, especially if scientific jargon can be introduced that does not usurp their burgeoning understanding.

The lead facilitator briefly summarizes that galaxies are made of stars, gas, dust, and any other astronomical object the learners correctly mentioned (e.g., black holes, planets, dark matter). Now the "starter" images are shared to demonstrate how galaxies are diverse in how they look. For example, with the big book of galaxies (e.g., Sandage & Bedke, 1998), the facilitator would page through to show different spiral galaxies. Alternatively, if a color image of many galaxies were used, the facilitator would point out a diverse selection of them. It is stated that galaxies are similar and different in their shapes and colors; a good analogy is how the people in the room are similar and different and how blood relatives share characteristics (also see §3): "we can often tell kids from adults based on heights" or "though all humans have noses, some relatives have noses much more similar to each other than a random person."

As seen in the galaxy images in Figure 1, there are other objects in the field of view (or even artifacts like bad pixels; see M109 thumbnail). It preempts some tangential questions and false classification criteria to note these foreground stars and background galaxies are not the main galaxy; a useful analogy is how a picture of a friend in public may have other random things like foreground bushes and background cars. As mentioned in §1 and further explored in §4, the galaxy-classification activity can lead to learning about observational astronomy so the facilitator may handle points about other objects and artifacts differently. Otherwise, a general facilitation trick is to defer or table tangents: "there's a lot to that observation; we'll try to get to it later, after we cover the basic points."

Classification is introduced as a general tool of science (e.g., animal kingdom), and objects in similar classes are described as reflecting shared physical properties, such as origins and/or evolution. The group is informed they will be scientists/astronomers classifying galaxies; they need to discuss why they think galaxies belong together or not, based on shape and color. It is explicitly stated that the galaxies are classified based on their observed properties and not on friends wanting to socialize. Making clear there will be a share-out helps the learners focus on justifying their classifications.

For younger learners, each student is given a letter-sized, color galaxy image (see Figure 3); the ensemble used by the class includes multiple examples of each galaxy type. Learners are told each galaxy is unique and has a great story, so they are not to trade. The instructions are to group sort, suggesting everyone walk around with their images facing outward. The facilitators mingle as well, helping[4] learners who seem unengaged or isolated by asking them what they think is important about their galaxy, then indicating other images that seem to share one or more key similarities, perhaps even facilitating the introduction to the other learner or group. It is not ideal for one learner to be a galaxy class of their own so, at the least, they should be situated near a group a facilitator considers related because it will help during the share-out and/or concluding remarks. The lead facilitator monitors the level of engagement and keeps time during the group-classification portion. When most groups seem formed, the learners are instructed[5] to sit with their group to indicate they are ready.

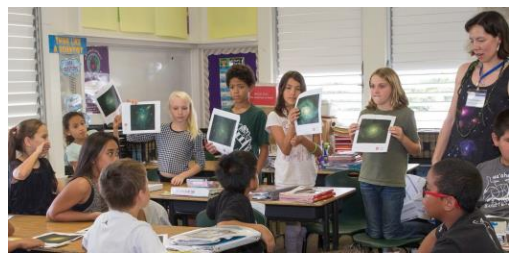

**Figure 3: Example of share-out by 5[th]-grade students.** In a class of about 30 students, these five determined their galaxies belonged together based on shape and color. In the concluding remarks, the lead facilitator (on the right) specifies these are face-on spiral galaxies and shares how they relate to other learner-defined groups of tilted or edge-on spiral galaxies.

### 2.2.2 Share-out

To begin the share-out, the lead facilitator asks a group to volunteer to explain why they decided their galaxies belong together and remind everyone to pay respectful attention. The group stands, holds their images where everyone can see them, and explains their classification (see Figure 3); the lead facilitator verifies they are done ("is there anything else to add?") before asking if anyone else in the room sees similarities. If anyone dissents, the facilitator acknowledges the input respectfully (e.g., "scientists do disagree sometimes"), notes how the point will be addressed later, and makes sure to return to it (usually in the concluding remarks). As mentioned previously, sometimes learners' criteria fixate on other objects or artifacts in the image (e.g., "all our galaxies have a bright red star in the corner"); the facilitator can address, defer, or table these points as suitable. Finally, the facilitator thanks the group for sharing and leads the applause. The process is repeated for all groups, with a final round of applause after everyone presents.

---

[4] A small yet impactful gesture is for the facilitator to squat, kneel, or sit to bring themselves eye-level with the learner. It contributes to the atmosphere of respect.

[5] In young classrooms, usually the teacher has a call-and-response trick to quickly attract the students' attention (e.g., instructor claps "shave and a haircut," then students clap "two bits" and focus quietly on the instructor); it is handy for a guest facilitator to learn this trick and use it!

### 2.2.3 Concluding remarks

The activity conclusion described here is very full and can easily take more time than remains. Thus, the LOs must be addressed early, and the concluding remarks should end with them, too. The lead facilitator must pay attention to the time; it is acceptable to stay light on details initially and dig deeper if time allows. For example, sometimes there is time to show the videos or demonstrate what learners may do next (e.g., contribute to the citizen-science project Galaxy Zoo). The conclusion is a balance of driving home the content-oriented LOs and honoring the learners' contributions, interests, and needs — the primary, motivation-oriented LOs.

The concluding remarks are an interactive narrative about the process and content of the galaxy-classification activity. The lead facilitator explicitly incorporates what the learners actually said and did, which contributes to feelings of accomplishment and ownership, and, especially for younger learners, maintains engagement and enthusiasm. However, as also emphasized in §3, the facilitators must discuss and plan the concluding remarks beforehand, so it is focused on the LOs.

While the lead facilitator uses the physical models (Figure 2) to illustrate the 3D geometry, 2D projection effects, etc., it is important to draw specific examples from the learners' images. The facilitator asks permission to borrow a learner's galaxy to make specific points to the whole group and thanks the learner when returning the image. "May I borrow your galaxy?" demonstrates respect and ownership; the "personal" story illustrated with the loaned galaxy also helps the learners be proud of their galaxies.

The lead facilitator highlights their predetermined LOs early in the concluding remarks while connecting the learners' contributions into how professional astronomers think about galaxy morphologies and what they tell us about galaxies more generally (see §1.1). It is important for the learners' engagement and enthusiasm to leave time for a general astronomy question-and-answer session; many people are just curious about black holes and aliens! The Q&A can be a time to tie up the deferred or even tabled points that arose.

The usual starting point is to connect the learner-identified galaxy classes together; as shown in Figure 3, often learners do not realize face-on spiral galaxies are related to tilted and edge-on spiral galaxies. The physical models (Figure 2) and, possibly, shadows are used to demonstrate projection effects; the effects of dust reddening will need to be addressed to "convince" learners the seemingly different colors are not inconsistent. It is useful to point out we observe the edge-on Milky Way disk from within, at night, where the obscuration of the diffuse light is the gas and dust within the Milky Way disk.

Often learners will have commented on the size, brightness, and color at the center of face-on spiral galaxies; this is an opening to talk about the central bulge, with its supermassive black hole and yellowish coloring due to its stellar population and reddening, using the donut-and-ball model (see Dunkin' Donuts frisbee and ball in Figure 2).

From there, the ball can be removed to begin the spiel about elliptical galaxies ranging from spheroidal to ellipsoidal (e.g., rugby ball), but, since they are round, they largely look the same from any angle; it is emphasized that their yellow-to-red color is due just to the older stellar population and not dust.

The lead facilitator segues by noting the ball from the spiral-galaxy model is the "wrong" size, since elliptical galaxies are formed from the mergers of smaller (often spiral) galaxies; the fact that the elliptical galaxies tend to be some of the largest galaxies helps learners who initially felt disappointed by their plain-looking galaxies. The facilitator can compare the spiral's yellow ball to the big yellow ball (see Figure 2) to visually emphasize the size difference or borrow two spiral- and one elliptical-galaxy images from learners to demonstrate a merger event.

Thus, the concluding remarks transition to galaxy evolution (see §1.1) and bring in the irregular galaxies, whether they have been highlighted in a learner-identified class, by solo students, and/or are lumped into other groups. As mentioned in §2.2.1, a good analogy for the process of classification is how the people in the room are similar and different. The expansion in the concluding remarks would include how people change over time; for example, "like how you grow up because you eat, galaxies also grow by 'eating' — other galaxies!" or "your hair can change color if it's dyed and then washes out."

The fact that irregular galaxies are ambiguous is (i) a true issue with visual classification and (ii) built into the activity with the choice of galaxy images; the suite of 33 images typically used has roughly 21 spiral, seven S0, and five elliptical galaxies with at least four spirals being reasonably classified as irregulars and several spiral and S0 galaxies showing merger signatures. At this point, the final handout is distributed so learners can see how "real" astronomers classify galaxies, but the bigger point is how the learners successfully recreated what professional astronomers did. This is a solid point to end on, especially when the primary LO is to foster interest in and motivation for pursuit of science.

## 2.3 Implementation for college learners

The galaxy-classification "starter" has been used for a college introductory astronomy course (roughly 20 students); the course was the second of a sequence, both required for astronomy majors. It covered stars, galaxies, and cosmology. The classification activity was placed at the start of the galaxy module, which came after a module on our Milky Way Galaxy. The students were previously taught properties of light, astronomical images, colors of stars of different masses and ages, and one spiral Galaxy; thus, the largest difference for this implementation of the classification activity is the expectation of what the learners know and can do.

For college learners, the activity is similar to the original COSMOS design (§2.1) but for multiple groups of learners (not just one small group). For share-out, each group writes their classification under each galaxy (called "share-out sheets"; see Figure 4).

It is assumed the lead facilitator is also the course instructor; if there were more than six groups, there should be more facilitators (e.g., teaching assistants).

Regarding timing, the classification activity is essentially a one-period "starter" (Kluger-Bell, 2010). One class period is 50–75 minutes, but, since the activity is part of a larger module, the concluding remarks can be 15 minutes or several subsequent class periods![6] With the standard suite of 33 galaxies, group discussion and classifying take up to 30 minutes, and this includes the share-out task of each group publicly documenting their classification for each galaxy (see Figure 4). Overall, five minutes may be sufficient to give the abbreviated instructions and set the groups in motion. It helps to have the assigned groups projected as students enter the classroom so they can self-organize.

---

[6] For example, Cooksey would have six 50-min follow-up lectures on galaxies and galaxy evolution, after the classification activity.

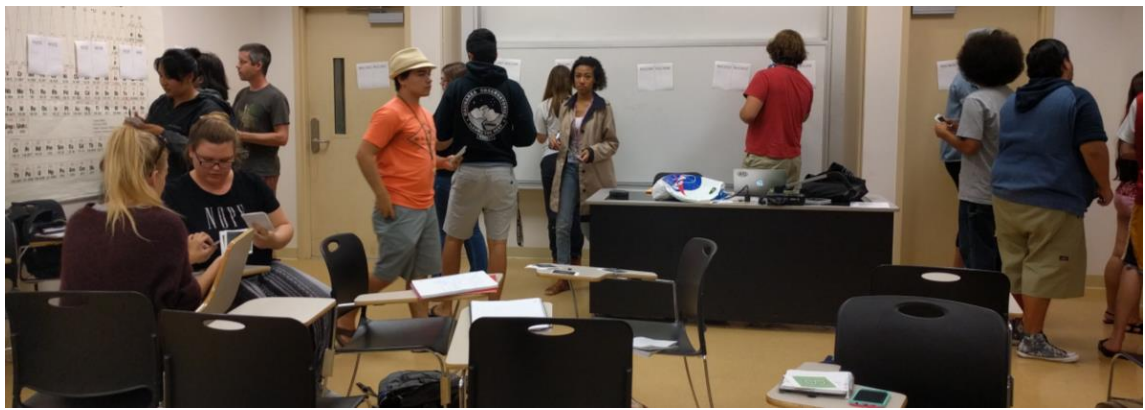

**Figure 4: College students documenting how their group classified each galaxy.** The share-out sheets are displayed around the room in alphabetical order of galaxy identifier, for convenience. When a group has classified all galaxies, they record their classification names under each galaxy. This share-out design leads itself to a very obvious timing for the activity and contributes to further learner discussion as they see what each other did.

### 2.3.1 Instructions and assigning groups

The basic instructions include: groups will be classifying a suite of galaxies, assigning a name for each classification; learners are to discuss their reasoning since "we are scientists"; when done, groups record their classification per galaxy on the share-out sheets arranged around the room (see Figure 4); and there will be concluding remarks afterwards. The lead facilitator makes it clear that a group may disagree internally; the group should record the classification with an "either/or" on the share-out sheets. The lead facilitator also states the facilitators will be mostly observing and available for questions at any time.

With the classification activity embedded in a larger course, the instructor can assign groups with attention to social dynamics and learners' prior knowledge and engagement; this is an essential facilitation task. From experience, an ideal group size is three (hard for anyone to hide); four works if the class size were not a multiple of three or if there were a need to keep the total number of groups manageable for the number of facilitators. Social dynamics refer to the learners engaging with each other in a way that furthers the learning process, so they need to actively contribute. The instructor should strategically place the typically engaged students, perhaps pairing them with disengaged or even recalcitrant peers. Groups should be assigned with attention to issues of diversity and inclusion in the sciences; for example, if possible, no group should have a minority of a demographic underrepresented in the sciences (e.g., a group of three has at least two female-identifying students).

Secondary to social dynamics are learners' prior knowledge and engagement because they affect peer learning. If grouping by similar prior knowledge, the lead facilitator should prepare an expansion or challenge prompt for groups who finish significantly ahead of others; this may be an extra set of more ambiguous galaxy images or a task to pinpoint and classify background galaxies (harder at lower resolution!) Another option is to assign each group a student from each tercile (or quartile) of the grades-to-date. However, the design of the classification activity deemphasizes the "correct" answer, so designing groups to have a range of prior knowledge mostly aims to deepen the discussions.

### 2.3.2 Active classification time and facilitation strategies

While groups are classifying, the facilitators primarily eavesdrop to aggregate points to leverage later (just as the facilitator honored learners' contributions in §2.2). If there has not been much prior group work, after about five minutes of activity time, the facilitators should make a round of quietly observing each group as they work so the learners get used to the facilitator being close-by without interrupting. Then the facilitators make a circuit and ask each group if they are doing all right or have questions. Otherwise, it is useful for the facilitators to observe and absorb what is happening. The facilitators monitor for any social issue and intervene as necessary (see Kluger-Bell et al., this volume); for example, if one group member is not participating, a facilitator can approach the group and ask the disengaged student directly what the group is considering. Particularly in this case (but also generally useful), the facilitators can squat, kneel, or sit to be eye-level with the students; it encourages engagement and shows respect.

As groups appear to be done, the lead facilitator reminds them to document their classifications for each galaxy on the share-out sheets. Once these are fairly populated, the facilitators review these in preparation for the concluding remarks. It is important to identify common names groups gave to different galaxy classes (at this level, learners likely know elliptical vs spiral vs irregular) and to mentally flag which galaxies generated significant consensus versus mixed results, for use in the concluding remarks.

### 2.3.3 Concluding remarks

In lieu of a formal share-out and as a transition from activity time to the conclusion, the lead facilitator asks about one of the notable mixed-result galaxies from the share-out sheets; for example, if a galaxy has one drastically different response (e.g., five spirals to one irregular), the facilitator asks the outlying group to share their rationale. The facilitator thanks the group and asks if anyone else in the room has a comment. Usually this opens the door to an informal discussion and naturally bleeds into the interactive concluding remarks to address the learners' points and the LOs.

The LOs for the classification activity are aligned with and embedded within the larger course objectives. Thus, what is covered in the activity's concluding remarks can be modulated by what the instructor knows will (or can) be covered in the upcoming class periods. It is good to refer to learner contributions from the classification activity as much as possible going forward in the course; this contributes to respect, ownership, engagement, and learning retention.

The concluding remarks begin as an interactive narrative with physical models (as detailed in §2.2.3; also see Figure 2) and may end with a formal presentation that introduces the "Hubble tuning fork" explicitly and reinforces the other LOs; we emphasize again: the facilitators must discuss and plan the concluding remarks beforehand, so it is focused on the LOs. If time allows, there may be an informal summative assessment where an image with many galaxies (e.g., *Hubble* Deep Field, Hickson Compact Group 44) is displayed and the class is asked what they observe and what they now understand about the objects.

## 3. Considerations for social and cultural aspects

Facilitation is crucial to the success of the classification activity in all its §2 forms, so facilitation strategies are interspersed in this article. They focus on respecting, engaging, assessing, and/or guiding the learners; they also support learner ownership. The suggestions range from bringing oneself to eye-level with the learner to prompts for various situations. When preparing to run the activity, the lead facilitator should make a facilitation crib sheet or bulleted notes in an instructor guide for use by all facilitators. The facilitators should also consider the social and cultural aspects of their learners and venue. Some social aspects may be logistics: if handling a very large number of learners, what modifications are necessary?; or, for a public-outreach event where learners will be brief participants, how to facilitate a quick galaxy-classification "playing card" activity?

A significant social consideration are learners with disabilities; a full discussion respecting the range of concerns and possible adaptations is beyond the scope of this article, and facilitators in need of guidance are recommended to turn to national and international professional organizations, such as the American Astronomical Society Working Group on Accessibility and Disability.[7] For example, to accommodate learners with color vision deficiency, the activity can be modified to use black-and-white images with a subsequent focus on morphology. For participants who are blind, engagement with astronomy can be accomplished with sounds and/or 3D models (Díaz-Merced, 2014).

Regarding both social and cultural considerations, the concluding remarks must be planned beforehand lest foot be stuck in mouth. It is certainly the case that learners can be forgiving of curious statements (to put it mildly); however, there is anecdotal evidence that even one bad exposure — especially when young — can be a life-long deterrent to science. Below we describe some social and cultural aspects as applied to running the activity on the Big Island of Hawai'i.

Local culture was incorporated into the language of the activity. For example, in exchange for galaxy "class" or "family," the term used was "galaxy 'ohana," since the idea of family is strong in Hawai'i. However, since 'ohana can include non-blood relations, examples need to be carefully chosen or phrased; even in §2.2.3, we carefully used "[blood] relatives" and not just "family."

Galaxies were defined as "islands of stars." This lent itself handily to a demonstration of size-scales in the Universe. If our Sun were a ball eight inches in diameter, the nearest stellar neighbor, Proxima Centauri, would be in the middle of mainland U.S (about 3600 miles away). However, if our Milky Way Galaxy were the size of the Big Island (about 75 miles wide, a scaling where the Sun would much be smaller than a grain of sand), the nearest galactic neighbors, the Magellanic Clouds, would be in west Maui, the next island in the chain (about 110 miles away). These facts were handy when discussing galaxy mergers since learners regularly asked what is going to happen to Earth when the Milky Way merges with the Andromeda Galaxy in 2.5 billion years: stars are highly unlikely to collide during a galaxy merger! (The real action is in the gas, dust, and dark matter; see Jonsson et al., 2008.)

# 4. Galaxy-classification activity as a stepping-stone

As discussed in §1, galaxies are a rich content area, so there are many avenues of investigation after learners grasp galaxy basics. For example, college students may be assigned a task with Galaxy Zoo (see §2.2).

Metevier et al. (2010) describe a community-college short course with a "research inquiry" on galaxy morphologies, normal and active galaxies (referring to their central supermassive black holes), and galaxy clusters (the largest gravitationally bound objects). If the whole Metevier et al. inquiry cannot be implemented, one could use the morphology section of that activity, which has LOs and activity elements that overlap with the galaxy-classification activity described here. One of the other inquiry topics (active galaxies, galaxy clusters) could be used or modified to deepen the "hands-on" or data-driven study of galaxies.

Similarly, Montgomery & Kulas (2010) outline a galaxy-component inquiry that focuses on the constituents of a spiral galaxy: stars; gas and dust; supermassive black hole; and dark matter. This provides a complete design to dig deeper on one galaxy type.

---

[7] See https://aas.org/comms/wgad.

As mentioned in §3, there is a pathway to quantitative analysis. In that section, a possible point in the concluding remarks was size scales in the Universe, with numbers given; instead, a follow-up activity or assignment would be facilitating the learners to make the calculations and inferences. Metevier et al. (2010) and Montgomery & Kulas (2010) also have quantitative elements described in their inquiry designs.

The classification activity could lead to an investigation of color, light, and spectra in astronomy. ISEE participants have developed such activities, though they are not publicly documented; their main LOs are: white light is composed of all colors, and the color(s) of light an object emits can help one learn about its temperature or chemical composition. A segue from galaxy taxonomy to properties of light may be to use cyan, magenta, yellow, and black transparencies of a galaxy; overlaying these one at a time on an overhead projector demonstrates subtractive color mixing. However, astronomical color images use additive color mixing, which can be demonstrated on a computer by overlaying the colored filter images of an object. The color images in Figure 1 are made from five filter images, colored to reproduce human vision.

Learners inspecting these images separately would see the structure of the galaxy and other objects in the image change because different components emit light in different colors. Only where the same object emits in all colors would an overlay result in white light; otherwise, different regions end up being predominantly e.g., blue or red, which, as covered in the classification-activity concluding remarks, means something astrophysically. Multi-filter images are a step toward understanding spectra, where a source of light is spread out in color space. A variant for a color, light, and spectra follow-up would be classifying galaxies imaged in different wavelengths, like infrared or ultraviolet; just like components may be bluer or redder in visible light for an astrophysical reason, so too would objects may appear different in infrared versus optical versus ultraviolet.

Thus, the galaxy-classification activity can be a stepping-stone not just to deeper understanding of galaxies but also other astronomy topics. LOs and logistics dictate how the classification activity is implemented and what comes after. Perhaps more importantly, successful experience with the widespread scientific practice of classification in an engaging and well-facilitated environment may motivate learners to pursue the study of science further.

# Acknowledgements

We would like to thank the support of Lisa Hunter and the education theme within the NSF CfAO (AST–9876783) which fostered the early development of the COSMOS astronomy course. KLC acknowledges partial support from NSF AST-1615296 and Gemini Observatory for running Journey through the Universe (http://www.gemini.edu/node/11817) where much "beta" testing occurred.

The PDP was a national program led by the UC Santa Cruz Institute for Scientist & Engineer Educators. The PDP was originally developed by the Center for Adaptive Optics with funding from the National Science Foundation (NSF) (PI: J. Nelson: AST#9876783), and was further developed with funding from the NSF (PI: L. Hunter: AST#0836053, DUE#0816754, DUE#1226140, AST#1347767, AST#1643390, AST#1743117) and University of California, Santa Cruz through funding to ISEE.

Images used in Figure 1 and the suite of galaxies offered by the corresponding author were culled from the Sloan Digital Sky Survey (SDSS) that provides modular education tools (e.g., http://voyages.sdss.org/expeditions/expedition-to-galaxies/galaxies-3/) and teacher notes (e.g., http://voyages.sdss.org/for-educators/ground-control/teacher-guides/teacher-guides-

expeditions/galaxies-teacher-notes/). Funding for the SDSS-III has been provided by the Alfred P. Sloan Foundation, the Participating Institutions, the National Science Foundation, and the U.S. Department of Energy Office of Science. The SDSS web site is www.sdss.org.

SDSS-III is managed by the Astrophysical Research Consortium for the Participating Institutions of the SDSS-III Collaboration including the University of Arizona, the Brazilian Participation Group, Brookhaven National Laboratory, Carnegie Mellon University, University of Florida, the French Participation Group, the German Participation Group, Harvard University, the Instituto de Astrofisica de Canarias, the Michigan State/Notre Dame/JINA Participation Group, Johns Hopkins University, Lawrence Berkeley National Laboratory, Max Planck Institute for Astrophysics, Max Planck Institute for Extraterrestrial Physics, New Mexico State University, New York University, Ohio State University, Pennsylvania State University, University of Portsmouth, Princeton University, the Spanish Participation Group, University of Tokyo, University of Utah, Vanderbilt University, University of Virginia, University of Washington, and Yale University.